\documentclass[12pt]{article}
\usepackage{amssymb,amsmath,graphicx,setspace,psfrag}
\psfrag{xratio}{$x/\ell_s$}
\psfrag{alphal}{$\alpha_\ell$}
\begin{document}
\title{Measuring (n,f) cross-sections of short-lived states}
\author{J. I. Katz \\ Department of Physics and McDonnell Center for the
Space Sciences \\ Washington University, St. Louis, Mo. 63130 \\ and \\
Lawrence Livermore National Laboratory, Livermore, Cal. 94550
\footnote{Proofs to: J. I. Katz, Dept. Physics, CB1105, Washington U.,
1 Brookings Dr., St. Louis, Mo. 63130; tel: 314-935-6202; facs:
314-935-6219}
\footnote{28 pp., no tables, 4 figs.}
\footnote{Email: \tt katz@wuphys.wustl.edu}}
\date{\today}
\maketitle
\newpage
\begin{abstract}
This paper reviews measurements of fission cross-sections of short-lived
nuclear states, summarizes the formidable experimental difficulties involved
and suggests novel methods of overcoming some of those difficulties.  It is
specifically concerned with the two such states that have been well
characterized, the $J^\pi = {1 \over 2}^+$ (26 m) isomeric $^{235m}$U and 
the $J^\pi = 1^-$ (16 h) ground state (shorter lived than the isomer)
$^{242gs}$Am, and with measuring their fission cross-sections at MeV
energies.  These measurements are formidably difficult, partly because of
the need to produce, separate and collect the short-lived states before
they decay, and partly because of their comparatively small fission
cross-sections at these energies.  I present quantitative calculations of
the efficiency of advection of recoiling $^{235m}$U isomers by flowing gas
in competition with diffusive loss to the surface containing the mother
$^{239}$Pu.  This paper reports the initial development and evaluation of
some of the methods that must be developed to make the experiments feasible.
\end{abstract}
\newpage
\section{Introduction}
In order to test theories \cite{yb03a,yb03b,ybb04,t07} of fission
cross-sections, it is desirable to
measure the (n,f) cross-sections of both an isomeric state and the ground
state of the same isotope of the same element.  This would test the
ability to calculate effects that depend only on spin and parity without
the confusion of effects that depend on (Z,A).  At present there is no 
isotope for which the (n,f) cross-section has been measured successfully
at MeV energies, where theoretical predictions can be made.  Measurements
\cite{mu84,t87,deer88} exist only at thermal and cold energies, where
the cross-section is expected to be dominated by incalculable resonances, so
that it may not be a useful test of theoretical predictions.

The purpose of this paper is to estimate the parameters of the sources of
the short-lived states and of their irradiation that would be required to
obtain useful data.  This is necessary to justify further investigation.
Because these experiments would be difficult and ambitious, and (unlike the
easier experiments with thermal neutrons) have never been attempted,
bringing these ideas to the attention of the experimental nuclear physics
and reactor communities will expose them to useful criticism and further
their development to the point at which detailed experimental design would
become appropriate.

There are two fissionable isotopes that have more than one level with a
lifetime long enough to offer some hope of performing the measurement: 
\begin{itemize}
\item $^{235}$U has a 77 eV ${1 \over 2}^+$ excited state of 26 m
half-life in addition to its long-lived (700,000,000 y) ${7 \over 2}^-$
ground state.
\item $^{242}$Am has a long-lived (141 y) $5^-$ isomer and a short-lived
(16 h) $1^-$ ground state; this isotope has the unusual (but not unique)
property that almost all of it is found in the isomeric state.
\end{itemize}

     In each case the chief experimental obstacle is the brief lifetime of
the shorter-lived state, making it difficult to accumulate a quantity
sufficient for experimental measurement.  In addition, direct separation of
isomers from ground states has never been demonstrated (isomeric shifts of
the hyperfine structure of atomic levels make it possible, in principle, by
the methods of atomic vapor laser isotope separation) so that the shorter
lived states must be separated at birth.  Fortunately, separation at birth
is possible.

The experimental and technical difficulties are formidable.  Most of them
have been considered before, but the earlier literature, which I cite
whenever possible, is incomplete because unpromising or unsuccessful work is
often not documented.  This paper includes a review of an ``underground'' of
work on these problems.  It also proposes and calculates as well as is
feasible with present understanding of the difficulties.  Yet, with sufficient
resources and determination, it may be possible to overcome these obstacles.

This paper is the full report of the work presented \cite{k09} at the
November, 2009 meeting of the American Nuclear Society in Washington, D.~C.

\section{$^{235m}$U}
\label{uranium}

$^{235m}$U is the product of the 5.245 MeV alpha-decay of $^{239}$Pu, with a
recoil energy of 89 KeV.  More than 99.9\% of the decays are to the isomer.
Hence $^{239}$Pu is a steady source of the uranium isomer.  However, because
of its rapid decay the isomer soon becomes a trace constituent of the
ground state $^{235}$U.  In order to do experiments on the isomer, it must
be separated from the plutonium source in a few tens of minutes or less.

One possible approach would be to chemically purify the plutonium, removing
the uranium that has built up, and then repurify, extracting the newly
created isomer perhaps 10 minutes later.  This is a formidable problem
in radiochemistry but has been solved on a small scale \cite{t87}.  The
first purification must completely extract all the uranium, which will have
been in the ground state (the tolerable fraction of the uranium remaining is
less than the ratio of the isomeric half life to the time since fabrication
of the sample, or $< 0.0006$ for a month-old sample), and the second
purification, taking no longer than about 10 minutes, must extract 2 ppb of
newly produced (isomeric) uranium from plutonium.  We estimate in
Section~\ref{target} that to measure the cross-section at MeV energies
will require isolation of about 100 times as many isomeric nuclei as
achieved by \cite{t87} for their measurements at thermal energies.

Another approach uses the recoil of the newly formed $^{235m}$U to separate
it from the plutonium.  This recoil permits the $^{235m}$U to penetrate a
small, and uncertain, thickness of parent metallic Pu or Pu compound.
Measurements \cite{cc83} on other $\alpha$-decay recoil nuclei suggest a
stopping column density $\approx 20 \mu\,$g/cm$^2$, and therefore a stopping
length $\ell \approx 0.01\,\mu$ in metallic plutonium.  The $^{235m}$U
escaping from the plutonium may be collected electrostatically
\cite{mu84,deer88} or swept up in a stream of gas \cite{tbs87}, possibly
containing aerosols to which it would be attracted and bound if ionized
\cite{deer88}, and collected. 

Previous attempts to collect $^{235m}$U from gas flowing through a tube
coated with Pu have not succeeded \cite{t87,s03}, perhaps because diffusion
in the carrier gas permitted the $^{235m}$U to deposit on unintended
surfaces, or because flow of the carrier gas swept it around the intended
collector.  Free atoms will stick to almost any surface they encounter.  The
first problem is acute at low gas densities (where the diffusivity is large)
and the second is acute at high gas densities (where the diffusivity is
small).

For the mother isotope of number density $n_m$ and half-life $t_m = $ 24,100
y, and the daughter with half-life $t_d = 26$ m, collecting the daughter for a
time $t$ produces a yield
\begin{equation}
Y_d = {1 \over 2} n_m \ell  {t_d \over t_m} \left(1 - 2^{(-t/t_d)}\right) =
1.5 \times 10^7 {\ell \over 10^{-6}\,{\rm cm}}\ {\rm cm^{-2}},
\end{equation}
where the leading factor of $1/2$ is the fraction of upward directed
daughters and we have taken $t = t_d$ and $n_m = 4 \times 10^{22}$ cm$^{-3}$
for $\delta$-plutonium.  The parenthesis then introduces another factor of
$1/2$; longer collection increases this factor slowly, but at the price of
growing, eventually dominant, contamination with the ground state.

To produce $N = 4 \times 10^{12}$ $^{235m}$U nuclei, as estimated in Section
\ref{target} for the irradiation target, would require a plutonium source of
area $30 \times (10^{-8}\,{\rm m} / \ell)$ m$^2$.  This may be feasible,
but is challenging; the source could consist of long porous tubes through
whose walls the carrier gas would flow.

In any method of producing $^{235m}$U speed is of the essence because of its
26 minute half-life.  Talbert, {\it et al.\/} \cite{t87} have shown that it
is possible to perform complex chemical extractions and purifications,
followed by sample preparation and irradiation, within these time limits.
Even when recoil is used to separate the $^{235m}$U from its mother
$^{239}$Pu, the isomer will be accompanied by some sputtered plutonium.
This can be removed chemically along with the separation of the isomer from
the material on which it is collected.

\subsection{Flow-through sources and collectors}

\begin{figure}[p]
\begin{center}
\includegraphics[width=5in,height=5in]{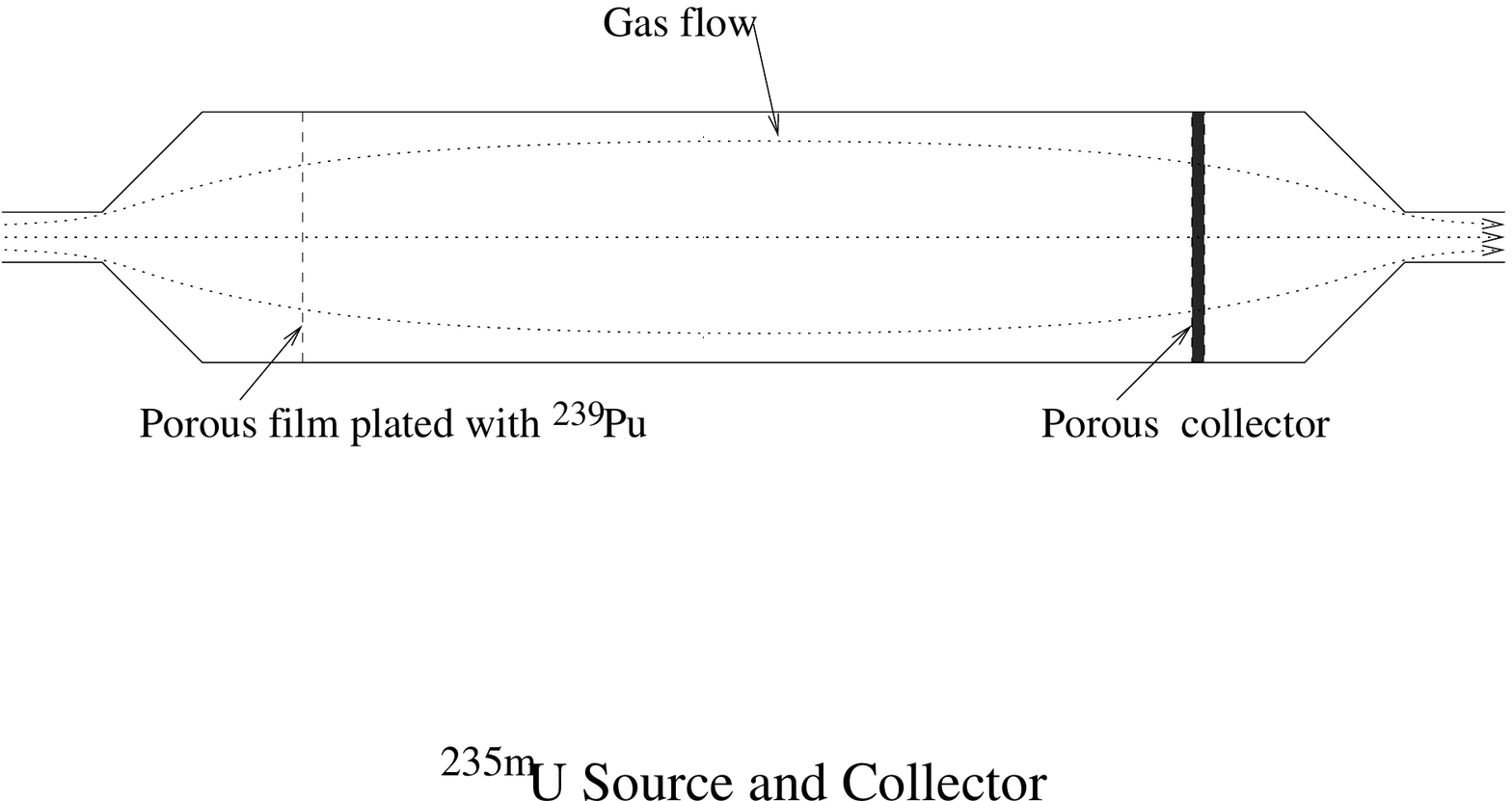}
\end{center}
\caption{}
\label{u235m}
\end{figure}

Rather than trying to find an optimum carrier gas density, we propose an
alternative solution, flowing gas through a porous source and a porous
collector.  The principle is that of a flyswatter: A porous object admits
the passage of gas, while trapping particles suspended in it.  The
microphysics is different, relying on diffusion of atomic species to solid
surfaces (the atomic species are smaller than the pores, unlike the fly).
For example, deposit the plutonium source on the downstream surface of a
nuclepore (polycarbonate) filter through which carrier gas will flow, being
careful that the pores remain unblocked.  
The configuration is illustrated in Figure~\ref{u235m}.

To determine feasibility, we first estimate the stopping length of
$^{235m}$U in the {\it gas\/}.  It loses about half its momentum each time
it passes through a column density of gas with mass equal to its own mass
divided by a collision cross-section.  Hence its stopping (thermalization)
length 
\begin{equation}
\ell_s \approx {\ln{(E_{recoil}/k_B T)} \over 2 \ln 2} {m_U \over \rho_{gas}
\sigma_{coll}} \approx 10 {m_U \over \rho_{gas} \sigma_{coll}},
\label{elleq}
\end{equation}
where $\rho_{gas}$ is the gas density, $\sigma_{coll}$ the effective
collision cross-section, which we take to be $10^{-19}$ m$^2$ ($10^{-15}$
cm$^2$), and room temperature gas is assumed.  The angle-averaged injection
distance is $\ell_s /2$.  This is generally quite short; $\rho_{gas} \ell_s
/2 \approx 2 \times 10^{-6}$ g/cm$^2$.


Collection is achieved by passing the gas through another filter or a
porous sacrificial medium.  The $^{235m}$U will deposit on its front
surface and the inside of its pores, and may be separated by chemical
processing of the entire collector.  If the pores are narrow the collector
need not be thick because of the rapid diffusion of the $^{235m}$U atoms
through the gas to the walls of the pores.  If they attach to aerosols in
the gas flow, the aerosols may be trapped by the pores.

\subsection{Collection efficiency}

We calculate the collection efficiency of $^{235m}$U injected into a carrier
gas flow from a porous surface through which the gas penetrates.  The
results are very generally applicable to any problem in which there is
competition between advection of some species from a volume source (in
this case, the volume is the region through which isomers recoiling from
the solid plutonium-containing surface are stopped in the gas) and diffusion
of that species to a surface to which they stick.

The fundamental equation is that of conservation of particles:
\begin{equation}
{\partial n \over \partial t} + V {\partial n \over \partial x} - D
{\partial^2 n \over \partial x^2} - S(x) = 0,
\end{equation}
where $n(x,t)$ is the density of the particles, we assume planar symmetry
and steady uniform gas flow at the speed $V$ in the $+ x$ direction, $D$ is
the diffusion coefficient of the particles in the gas and $S(x)$ is their
source strength.  We consider only stationary solutions, and replace partial
derivatives with respect to $x$ by ordinary derivatives:
\begin{equation}
V {dn \over dx} - D {d^2 n \over dx^2} - S(x) = 0.
\label{stateq}
\end{equation}

For an isotropic source of particles with stopping length $\ell_s$ the
volume source is readily seen to be uniformly distributed in $x$:
\begin{equation}
S(x) = 
\begin{cases}
S_0/(2 \ell_s) & x \le \ell_s \\
0 & x > \ell_s,
\end{cases}
\end{equation}
where $S_0$ is the areal source strength (particles per unit area per
unit time) and the factor of 2 comes from the fact that half the particles
are emitted in the $- x$ direction and do not enter the carrier gas.

Writing $y \equiv dn/dx$ for convenience, the only physically meaningful
solution for $x > \ell_s$ where $S(x) = 0$ is $y = 0$; any other solution
would diverge exponentially as $x \to \infty$.  Then for $x > \ell_s$ the
density of isomers $n(x)$ is constant:
\begin{equation}
n(x) = n_\infty,
\end{equation}
where the flux (per unit area) is $n_\infty V$.

For $0 \le x \le \ell_s$ Eq.~\ref{stateq} becomes,
\begin{equation}
{d y \over dx} = {V \over D} y - {S(x) \over D}.
\end{equation}
This is readily integrated:
\begin{equation}
y = C \exp{(Vx/D)} - {S_0 \over 2 \ell_s V} \exp{(Vx/D)}
\left(1-\exp{(-Vx/D)}\right).
\end{equation}
The constant of integration $C$ is determined by requiring continuity of
$y$ at $x = \ell_s$ where $y = 0$:
\begin{equation}
C = - {S_0 \over 2 \ell_s V} \left(\exp{(-\alpha_\ell)}-1\right),
\end{equation}
where we have defined the parameter $\alpha_\ell \equiv V \ell_s /D$.
Then for $0 \le x \le \ell_s$
\begin{equation}
y = - {S_0 \over 2 \ell_s V} \left(\exp{(V(x-\ell_s)/D)} - 1\right).
\label{yeq}
\end{equation}

Eq.~\ref{yeq} is readily integrated to give $n(x)$ for $0 \le x \le \ell_s$:
\begin{equation}
n(x) = - {S_0 D \over 2 \ell_s V^2} \exp{(-\alpha_\ell)} \left(\exp{(Vx/D)}
- 1\right) + {S_0 x \over 2 \ell_s V},
\label{nsoln}
\end{equation}
where we have used the boundary condition, appropriate to a surface that is
a perfect sink (any isomers diffusing to it stick to it, as will generally
be true of free atoms or ions) $n(0) = 0$.  Defining $\alpha \equiv Vx/D$,
and $n_0 \equiv S_0/2V$ (the isomer density in the carrier gas if there
were no losses by diffusion to the surface), Eq.~\ref{nsoln} can be rewritten
\begin{equation}
n(x) = {n_0 \over \alpha_\ell} \left[\alpha - \exp{(-\alpha_\ell)}\left(
\exp{(\alpha)} - 1\right) \right].
\label{neq}
\end{equation}

For $x > \ell_s$ $n(x) = n_{\infty}$, where $n_\infty$ may be found by
continuity, taking $x = \ell_s$ ($\alpha = \alpha_\ell$) in Eq.~\ref{neq}:
\begin{equation}
n_\infty = n_0 \left(1 - {1 - \exp{(-\alpha_\ell)} \over \alpha_\ell}
\right).
\end{equation}
This has the limits
\begin{equation}
{n_\infty \over n_0} =
\begin{cases}
{\alpha_\ell \over 2} - {\alpha_\ell^2 \over 6} + \cdots & \alpha_\ell
\to 0 \\
1 - {1 \over \alpha_\ell} + \cdots & \alpha_\ell \to \infty.
\end{cases}
\end{equation}

\begin{figure}[p]
\begin{center}
\includegraphics[width=5in,height=7in]{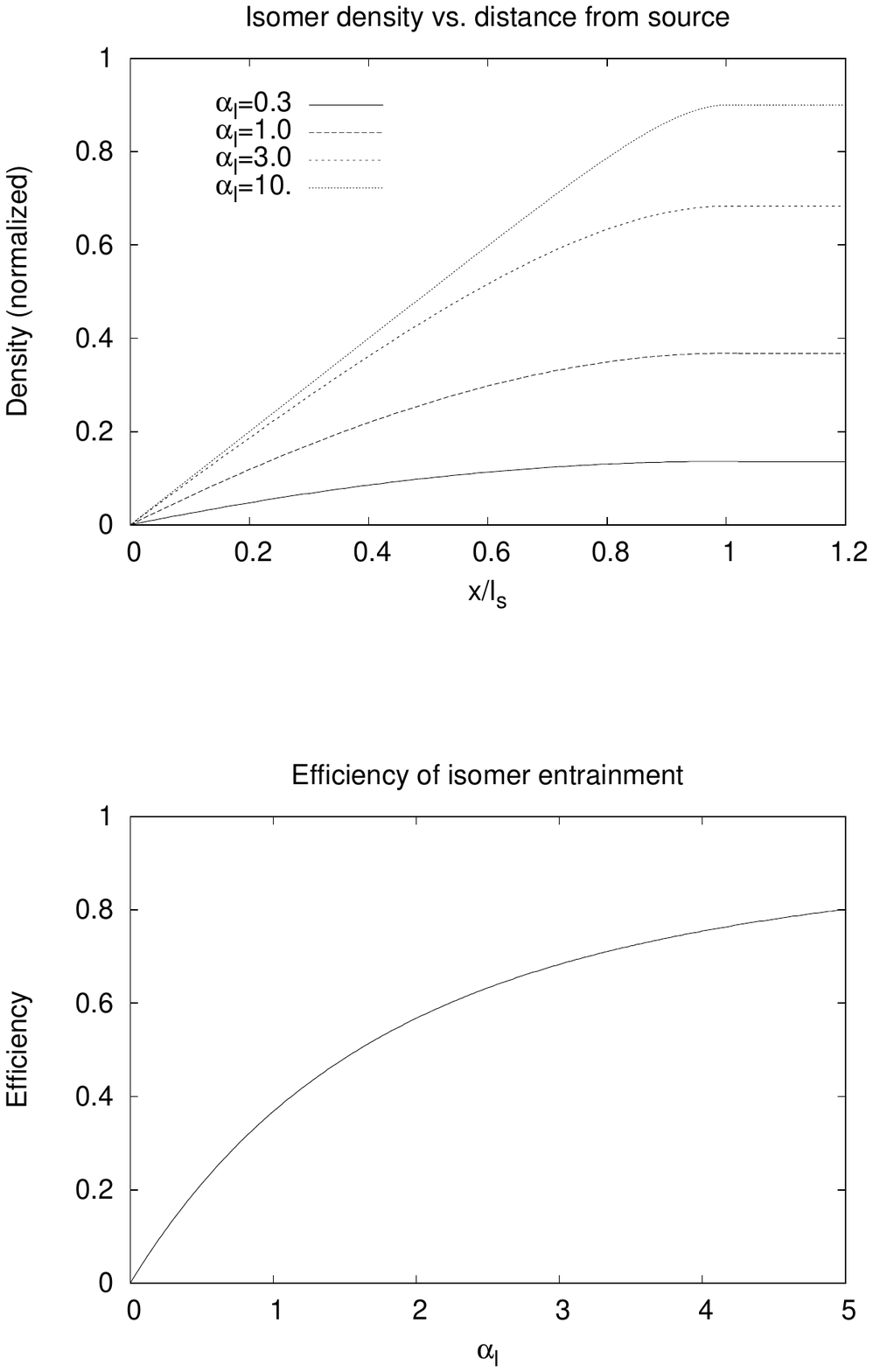}
\end{center}
\caption{}
\label{efffig}
\end{figure}

The fraction of the recoiling $^{235m}$U that are swept into the gas flow
is the collection efficiency $\epsilon \equiv n_\infty / n_0$.  The
remaining isomers diffuse back to the source and stick to its surface.
Using our previous estimate \ref{elleq} for $\ell_s$ and approximating $D
\approx v_{thU} m_U /(3 \rho_{gas} \sigma_{coll})$, where $v_{thU}$ is the
thermal velocity of a uranium atom, we obtain
\begin{equation}
\alpha_\ell \approx 30 {V \over v_{thU}}.
\end{equation}
These results are summarized in Fig.~\ref{efffig}.  For example, the
condition $\alpha_\ell > 2$ ($\epsilon > 0.57$) corresponds, at room
temperature, to $V > 7$ m/s.  This is not difficult to provide, and implies
a very low Mach number and little dissipation in the gas flow.

\section{$^{242gs}$Am}

\begin{figure}[p]
\begin{center}
\includegraphics[width=5in]{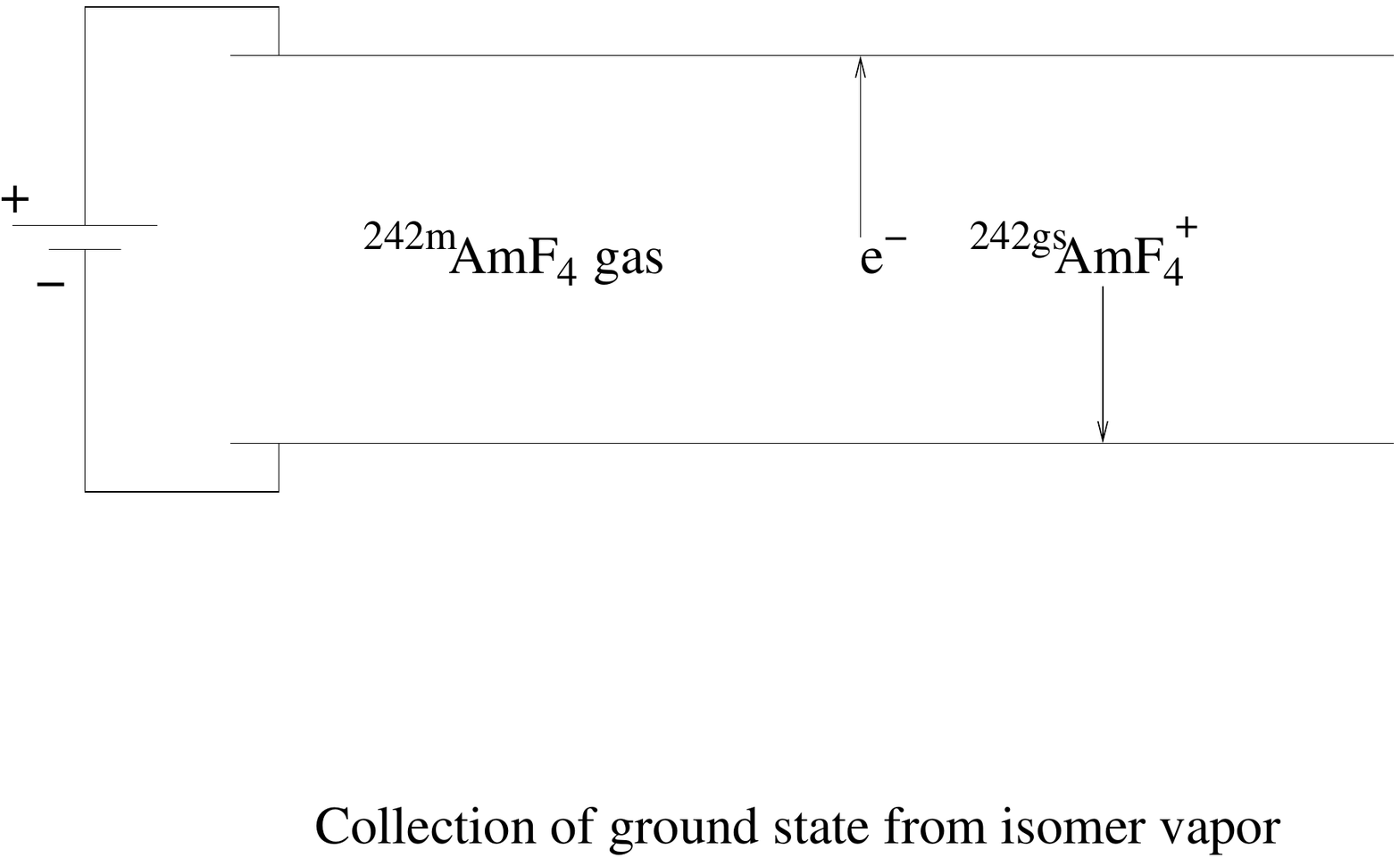}
\end{center}
\caption{}
\label{am242gs}
\end{figure}

$^{242gs}$Am is produced by internal transition of $^{242m}$Am.  At least
one electron (usually from the L shell) is ejected, ionized by the 49 KeV of
the excited state, and more may be ejected as the hole is filled by a
cascade.  Hence the ground state product is ionized and may be collected
electrostatically, just as for $^{235m}$U \cite{mu84,deer88}, as shown in
Figure \ref{am242gs}.  Subsequent isolation of the short-lived ground state
requires only chemical separation from the collecting electrode and from its
(chemically distinct) decay products.

Suppose we wish to collect $N$ $^{242gs}$Am nuclei from a region of area $A$
between two capacitor plates, separated by a distance $d$, in a time $t = 
t_{gs\,1/2} / \ln{2}$, after which 63\% of the asymptotic limit has been
collected (longer collections contaminate the collecting cathode with
fissionable $^{242}$Pu and $^{242}$Cm, although these can be removed
chemically when the $^{242gs}$Am is extracted, unlike the otherwise
analogous contamination of $^{235m}$U by $^{235gs}$U).  The volume $A d$ is
filled with $^{242m}$Am, as some suitable vapor at number density $n_v$, and
decays at a rate $\nu_m$.
The density and separation are limited by charge exchange, with cross-section
$\sigma_{ce}$, between the ionized $^{242gs}$Am-containing species and the
neutral $^{242m}$Am species, so that $n_v d \lesssim 1/\sigma_{ce}$, with a
strong inequality desirable.  The ground state production rate must be at
least $N/t$, from which we obtain the minimum area
\begin{equation}
A > {N \over t \nu_m n_v d} > {N \sigma_{ce} \over t \nu_m} = N \sigma_{ce}
{t_{m\,1/2} \over t_{gs\,1/2}} \approx 80 {N \over 10^{11}}\,{\rm cm}^2,
\label{amarea}
\end{equation}
where we have taken $\sigma_{ce} = {\cal O}(10^{-18}$ m$^2$) (${\cal O}
(10^{-14}$ cm$^2$)), generally a
conservative upper bound to atomic and molecular resonant $\sigma_{ce}$.
The required $A$ is feasible if $N \lesssim 10^{12}$, as estimated in
Section \ref{target}.

The assumed $n_v d < 10^{18}$/m$^2$ ($10^{14}$/cm$^2$) suggests densities
$n_v = 10^{19}$--$10^{20}$/m$^3$ ($10^{13}$--$10^{14}$/cm$^3$) in order to
keep the electrode separation in the convenient range 1--10 cm.  At
temperatures ${\cal O} (10^3)^{\,\circ}$K this range corresponds to
pressures ${\cal O}$ (0.1--1 Pa) (1--10 dyne/cm$^2$, $10^{-3}$--$10^{-2}$
mm-Hg).  

Americium fluorides are not volatile (unlike UF$_6$).  The most volatile
fluoride, AmF$_4$ \cite{cc70}, disproportionates (AmF$_4$ $\to$ AmF$_3 + {1
\over 2}$ F$_2$) above about 900$^{\,\circ}$K, at which $n_v \approx 2
\times 10^{18}$/m$^3$ ($p \approx 10^{-4}$ mm-Hg) \cite{gh87}.  AmF$_4$
vapor is very corrosive because of its readiness to fluorinate even
non-reactive metals.

AmF$_3$ is significantly less volatile than AmF$_4$.  A vapor pressure of
$10^{-2}$ mm-Hg requires $T \approx 1500^{\,\circ}$K \cite{cc55,k94}, while
even $10^{-4}$ mm-Hg, obtainable with AmF$_4$, requires $T \approx
1300^{\,\circ}$K.  For metallic americium the corresponding temperatures are
about 1260$^{\,\circ}$K and 1060$^{\,\circ}$K, respectively, lower than
those required for the trifluoride \cite{wkh79}.  In addition, it has no
``chemistry'' to deal with.

It might be possible to use tetrafluoride vapor, but a larger gas volume
$Ad$ would be required than for higher gas densities.  If $N$ is as small
as ${\cal O} (10^{11})$, as suggested in Section \ref{target}, than 
the lower vapor densities (than the upper limits permitted by charge
exchange) permitted by the AmF$_4$ disproportionation temperature limit
may be acceptable, while keeping the total source volume within bounds.
Then
\begin{equation}
Ad \gtrsim {N \over n_v}{t_m \over t_{gs}} \approx {N \over 10^{11}}
{10^{18}\,{\rm m}^{-3} \over n_v}\ 8 \times 10^{-3}\ {\rm m}^3.
\label{amvol}
\end{equation}
This appears feasible, but depends on the chemical stability of AmF$_4$
vapor over times of several hours at 900$^{\,\circ}$K.  The vapor could
be replenished continually, and later reconstituted from its
disproportionation products, but this would add an additional complication,
and raise the danger of loss of scarce material.

Americium organometallic compounds and oxides are much more promising 
because of their greater volatility \cite{d70,t85,d02,d10}.  Quantitative
measurements do not appear to exist, but temperatures $\approx 
130^{\,\circ}$C may provide sufficient vapor density.

The source dimensions scale according to the two constraints:
\begin{enumerate}
\item Loss of $^{242gs}$Am sets an upper bound $d \lesssim 10^{14}\,{\rm
cm}^{-2}/n_v$.
\item Providing sufficient $^{242gs}$Am sets the lower bound (\ref{amvol})
on the source volume.  Higher vapor density $n_v$ (requiring higher
temperature) would permit smaller vapor volumes, but would not reduce the
minimum collecting area (\ref{amarea}).
\end{enumerate}
These constraints are essentially independent of the vapor species, but the
required temperature depends, of course, on the equilibrium vapor pressure.

\section{The Neutron Source}

\begin{figure}[p]
\begin{center}
\includegraphics[width=5in,angle=270]{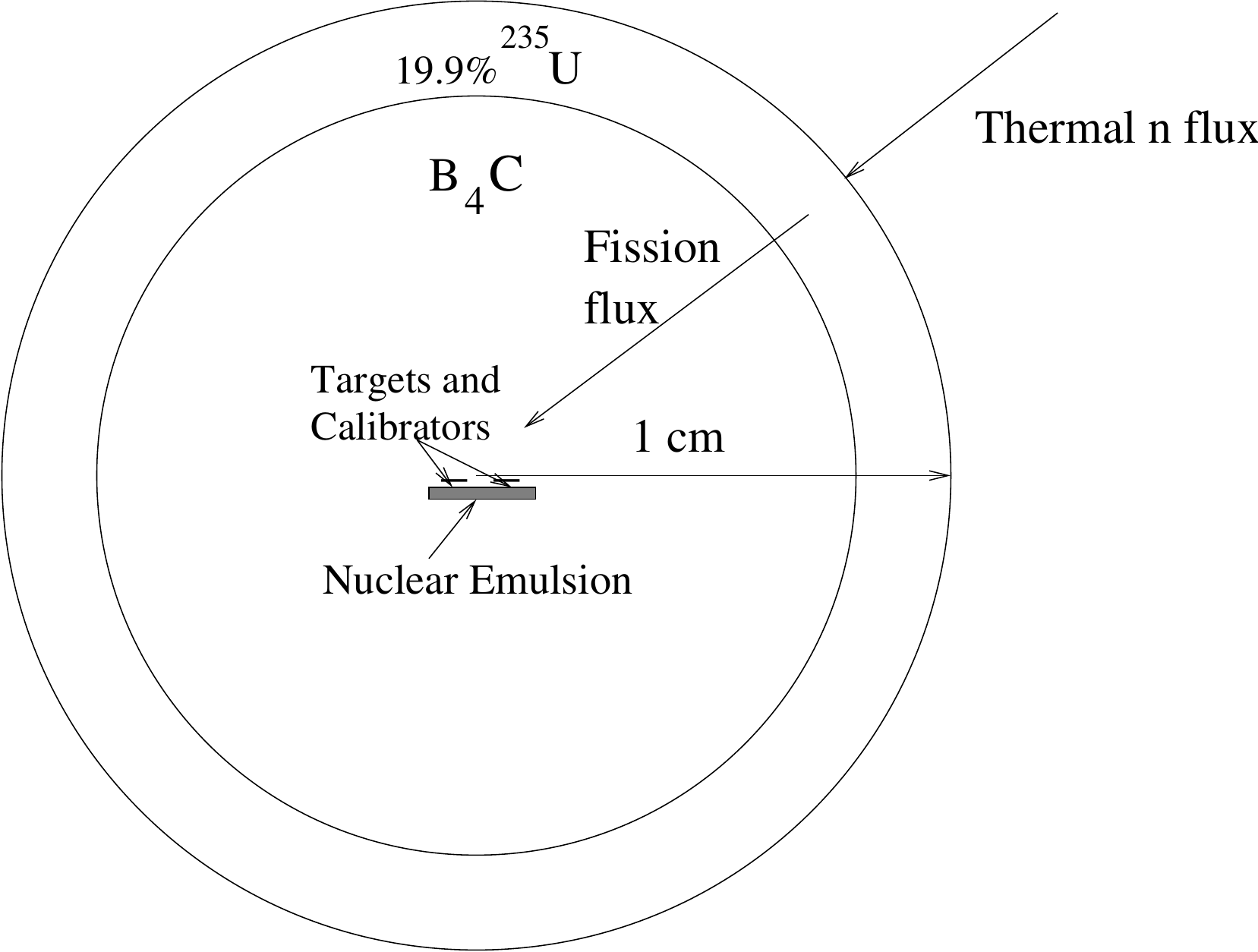}
\end{center}
\caption{}
\label{nconverter}
\end{figure}

     Measurement of the (n,f) cross-section at MeV energies requires an
intense source of MeV neutrons.  This may be obtained by conversion of the
thermal neutron flux of a reactor.  Incident upon a shell of 19.9\% enriched
$^{235}$U of thickness $\delta$, as shown in Figure \ref{nconverter}, a
fraction (independent of the shell radius as long as it is $\gg \delta$)
\begin{equation}
f_{fiss} = 1 - E_2(n_{235} \delta \sigma_{th\,(n,f)\,235})
\end{equation}
of the thermal flux produces fissions in 19.9\% enriched uranium, where
$n_{235} = 9.6 \times 10^{27}$ m$^{-3}$ ($9.6 \times 10^{21}$ cm$^{-3}$),
$\sigma_{th\,(n,f)\,235} = 580$ b is
its thermal neutron fission cross-section and $E_2$ is the exponential
integral.  For $\delta = 0.2$ cm the argument of $E_2$ is 1.12, $f_{fiss} =
0.87$ (most of the thermal neutrons are incident obliquely and have much
longer paths in the shell than at normal incidence) and the fission neutron
flux is $\nu f_{fiss} \approx 2.0$ times the incident thermal neutron flux
where $\nu$ is the thermally induced fission neutron multiplicity.


The attenuation of the fission neutrons in the uranium shell is readily
estimated.  The scattering probability (the cross-section is about 7 b
on both uranium isotopes) is 0.07 on a radial path.  Integrating over the
isotropic source spectrum yields a transmission probability of 0.78.  For
the illustrated geometry we are interested only in nearly radial fission
neutrons (because the target to be irradiated is near the center of the
sphere), and their transmission probability is 0.93.

The residual thermal neutron flux must be absorbed so that it induces no
more than a few percent as many fissions as the fast fission neutrons.
This is readily accomplished by filling the inside of the enriched
uranium shell with boron or boron carbide, possibly enriched with $^{10}$B.
A thickness of 8 mm of natural (20\% $^{10}$B) B$_4$C, pressed to solid
density ($\rho = 2.52 \times 10^3$ kg/m$^3$, 2.52 gm/cm$^3$), provides 54
$e$-folds of attenuation.

Ten $e$-folds are sufficient to reduce the thermal flux to levels at which
it induces, for plausible assumptions as to the cross-sections, only a few
percent as many fissions in the targets as the fission-spectrum neutrons of
interest.  $^{10}$B, unlike the resonance absorbers Cd and Gd, is also
effective in attenuating the epithermal neutrons present in reactors.

Attenuation of the desired fission (MeV) neutrons by the B$_4$C is readily
estimated.  Their total cross-section with either boron isotope or carbon is
about 2 b, and the fraction surviving the 8 mm radial path is about 80\%.

The achievable neutron flux is limited, not by the capabilities of research
reactors, but by the heat load of the fissions in the enriched uranium
shell.  A thermal neutron flux of $10^{16}$/(m$^2$\,s)
($10^{12}$/(cm$^2$\,s)) produces a fission power of 250,000 W/m$^2$ (25
W/cm$^2$).  This is a few times the heat flux of a kitchen
stove on its high setting, and can be dissipated by free convection if
immersed in water, or by forced gas flow, but represents a rough upper limit
to the tolerable flux.   

The thermal neutron calculations in this section are necessarily approximate
because they do not allow for the evolution of the neutron spectrum as it
is attenuated by processes ($^{10}$Be(n,$\alpha$)$^7$Li and
$^{235}$U(n,fiss)) whose cross-sections are energy dependent (generally
$\propto E^{-1/2}$ at thermal energies).  More accurate results, as well as
inclusion of the epithermal neutrons present in reactors, require the 
straightforward use of a Monte Carlo code such as MCNP.
\section{Required Target Quantities}
\label{target}
If $N_t$ target nuclei with fission cross-section $\sigma_{n,fiss}$, that
decay with half-life $t_{1/2}$, are irradiated by a neutron flux $F$ for a
time $t$, the number of fissions will be
\begin{equation}
N_{fiss} = N_t F \sigma_{n,fiss} \left(1 - \exp{(-t \ln{2}/t_{1/2})}\right).
\end{equation}
Although it might seem advantageous to take $t \gtrsim t_{1/2}$, that would
confound the fissions of the intended targets with those of their decay
products ($^{235}$U for $^{235m}$U and $^{242}$Pu and $^{242}$Cm for
$^{242gs}$Am) that are themselves fissionable.  The decay products have
cross-sections comparable to those of the intended targets, but that are
imperfectly known so that accurate subtraction would not be possible.  Hence
it is necessary to take $t \ll t_{1/2}$, and here we adopt $t = 0.1
t_{1/2}/\ln{2}$.

In order to limit the contribution of counting statistics to the error in
the measured cross-section to 3\%, $N_f = 1000$ would be required, noting
that each fission produces two oppositely directed tracks; for a target
plated on nuclear emulsion the counting efficiency may approach unity.  Then
\begin{equation}
N_t = 7 \times 10^{15} {N_f \over 1000}{\sigma_{n,fiss} \over 1\,{\rm b}}
{10^{12}\,{\rm cm^{-2}\,s^{-1}} \over F}{1\,{\rm s} \over t_{1/2}}.
\end{equation}
We use 1 b as a nominal cross-section for estimating the required target
mass.  Predicted (theoretical) isomeric fission cross-sections 
\cite{yb03a,yb03b,ybb04,t07} are typically about this value for MeV 
neutrons, so it sets a realistic scale for the required target masses.
Substituting the appropriate half lives and taking the fractions to be unity
yields
\begin{equation}
N_t =
\begin{cases} 4 \times 10^{12} & ^{235m}{\rm U} \\
1.2 \times 10^{11} & ^{242gs}{\rm Am}.
\end{cases}
\end{equation}

If the targets are plated as a monolayer, the areas required are each less
than 1 mm$^2$, and several targets can be placed near the center of the
neutron source, as illustrated.  Of course, thicker targets can be used,
with a corresponding reduction in area, so long as they are thin compared to
the fission product stopping length in high-Z material of several $\mu$.

The quantity of source material required is
\begin{equation}
N_{source} = N_t {t_{1/2\,source} \over t_{1/2}} =
\begin{cases} 2 \times 10^{21}\ (0.8\,{\rm g}) & ^{239}{\rm Pu} \\
9 \times 10^{15}\ (4\,\mu{\rm g}) & ^{242m}{\rm Am}.
\end{cases}
\end{equation}
Although the availability of $^{242m}$Am is limited, the quantity required
is very small; $^{239}$Pu is, of course, available in essentially unlimited
quantity.  
\section{Counting the Fissions}
     If the targets are plated upon nuclear emulsions, fissions occurring
during irradiation may be measured at leisure by counting etched tracks.
This has the advantage that counting is done in a low-radiation environment,
with an efficiency approaching 100\%, although it is painstaking work
through a microscope.  D'Eer, {\it et al.\/} \cite{deer88} counted
fissions during irradiation by coincident detection of fragments by two
semiconductor detectors, with the disadvantage of working in a very high
radiation environment.  In either case, a statistically 3\% accurate
cross-section would require counting only 1000 fissions.

Fissions are high energy events that produce a characteristic signature. 
Their energetic but low speed products deposit energy at a uniquely high
rate per unit length of their path; there is essentially no background.  In
a nuclear emulsion their tracks are unmistakable.   A particular advantage
is that multiple targets, both signal and calibration, can be irradiated
simultaneously and in close proximity (less than a mm apart) to ensure that
they are exposed to the same neutron flux, but with no ambiguity as to which
track is associated with which target because the tracks enter the emulsion
immediately under the targets.
\section{Calibration}
The proposed experiments would need calibration of the thermal neutron flux
at the irradiated targets.  This could be provided by a control sample of
$^{235}$U, and would amount to normalizing the measured cross-sections
to the (well known) thermal neutron fission cross-section of $^{235}$U.  The
flux of MeV neutrons would be monitored (it is intended to be very small) by
a control sample of $^{238}$U or $^{232}$Th with well known fission
cross-sections at that energy but essentially zero cross-section at thermal
energy.  In the test samples differential measurements (using different
periods of decay of the short-lived states) would be required to allow for
the effects of isotopic or isomeric impurity.

The quantities of the irradiated targets would be determined after their
natural decay.  The mass of $^{235m}$U could be determined by measuring the
fission rate of its daughter $^{235}$U in an independently calibrated
thermal neutron flux.  The mass of $^{242gs}$Am could be determined by
counting the 44 KeV (0.03\%) gamma rays or the $\alpha$ decays of its
(82.7\% branching ratio, 163 d) daughter $^{242}$Cm following chemical
separation.
\section*{Acknowledgments}
I thank B.~Beck, P.~J.~Bedrossian, L.~A.~Bernstein, J.~T.~Burke,
M.~B.~Chadwick, R.~J.~Fortner, F.~R.~Graziani, R.~C.~Haight,
R.~A.~Henderson, S.~B.~Libby, T.~C.~Luu, D.~P.~McNabb, K.~J.~Moody,
D.~A.~Shaughnessy, L.~G.~Sobotka, M.~A.~Stoyer, P.~Talou, D.~J.~Vieira and
J.~B.~Wilhelmy and anonymous referees for useful discussions, constructive
criticism and insight, and for calling my attention to volatile compounds of
americium, and the Lawrence Livermore National Laboratory for hospitality.

\newpage
\newpage
\newpage
\newpage
\newpage
\section{Figure Captions}
\begin{enumerate}
\item Schematic of $^{235m}$U source.  Inert gas flows through the
tube from left to right, passing through pores in the source and collector.
The source has a thin layer of $^{239}$Pu plated on a supporting film, with
gas flowing through it fast enough (see text) that $^{235m}$U that stop in
the gas do not diffuse back to the film.  The $^{235m}$U is swept by the gas
flow to the porous collector where it diffuses to and attaches to the
surface or to the walls of its pores, for later chemical extraction.
\item Upper graph shows the density of $^{235m}$U isomers as a function of
distance from a porous source through which gas flows at a speed $V$.  The
distance is normalized to the recoil stopping length in the gas $\ell_s$ and
the density normalized to $S_0/2V$, where $S_0$ is the source strength; this
would be the density were there no diffusive losses of isomers to the
surface of the source.  Lower graph shows the efficiency of entrainment of
isomers in the gas flow as a function of the parameter $\alpha_\ell \equiv V
\ell_s / D$, where $D$ is the diffusivity of isomers in the gas; this
describes the comparative rates of advection of isomers from their source to
that at which they diffuse back to it, the source surface acting as a sink.
\item Apparatus for electrostatic collection of $^{242gs}$Am from vapor.
The vapor is indicated as AmF$_4$, but it could be atomic Am or AmF$_3$.
\item Thermal to fission spectrum neutron converter, approximately to scale.
\end{enumerate}
\end{document}